\documentclass[12pt,a4paper,final]{iopart}
\expandafter\let\csname equation*\endcsname\relax
\expandafter\let\csname endequation*\endcsname\relax
\usepackage{iopams}
\usepackage{graphicx}
\usepackage{subfigure}
\usepackage[breaklinks=true,colorlinks=true,linkcolor=blue,urlcolor=blue,citecolor=blue]{hyperref}
\usepackage{cite}
\usepackage{amsopn}
\usepackage{braket}
\usepackage{mathtools}

\usepackage[a4paper]{geometry}
\usepackage{bpchem,upgreek}
	\usepackage{amsmath}
	\usepackage{makeidx}
	\usepackage{amsfonts}
	\usepackage[ansinew]{inputenc}
	\usepackage[usenames,dvipsnames]{pstricks}
    \usepackage{graphicx}
    \usepackage{float}
    \usepackage{subfigure}
	\usepackage{pst-grad} 
	\usepackage{pst-plot} 
	\usepackage{sidecap}
	\usepackage[makeroom]{cancel}

\begin{document}

\title[Quantum repeater protocol...]{Quantum repeater protocol in mixed single- and two-mode Tavis-Cummings models}

\author{M Ghasemi$^{1}$}
\author{M K Tavassoly$^{1}$}
\address{$^1$Atomic and Molecular Group, Faculty of Physics, Yazd University, Yazd  89195-741, Iran}
\ead{mktavassoly@yazd.ac.ir}

\vspace{10pt}
\date{today}

\begin{abstract}
In this paper we study the production of entanglement between two atoms which are far from each other. We consider a system including eight two-level atoms $(1,2,\cdots,8)$ such that any atom with its adjacent atom is in atomic Bell state, so that we have four separate pairs of maximally entangled states $(i,i+1)$ where $i=1,3,5,7$. Our purpose is to produce entanglement between the atomic pair (1, 8), while these two distant atoms have no interaction. By performing the interaction between adjacent non-entangled atomic pairs (2, 3) as well as (6, 7), each pair with a two-mode quantized field, the entanglement is produced between atoms (1, 4) and (5, 8), respectively. Finally, by applying an appropriate Bell state measurement (BSM) on atoms (4, 5) or performing an interaction between them with a single-mode field (quantum electrodynamic: QED method), the qubit pair $(1,8)$ becomes entangled and so the quantum repeater is successfully achieved. This swapped entanglement is then quantified via concurrence measure and the effects of coupling coefficients and detuning on the concurrence and success probability are numerically investigated. The maxima of concurrence and success probability and the corresponding time periods have been decreased by increasing the detuning in asymmetric condition in BSM method. Also, the effects of detuning, initial interaction time and coupling coefficient on the produced entanglement by QED method are considered. Increasing (decreasing) of the detuning (interaction time) has destructive effect on the swapped entanglement in asymmetric condition.
\end{abstract}


\vspace{2pc}
%
%
%
%

\section{Introduction}
Extending the quantum phenomena to macroscopic distances is of considerable interest \cite{Qin2017,Briegel1998}. In particular, entangled quantum states can be transferred over long distances using quantum repeater protocols \cite{Briegel1998}. In fact, to prevail the entanglement attenuation in the process of transferring the entanglement the quantum repeater is proposed. These transferred states are useful in quantum communication \cite{Li2016}, quantum key distribution \cite{Rubenok2013} and quantum cryptography \cite{muller1996}. In Ref. \cite{Su2018} the authors have investigated the optimal probability of creating a maximally
entangled state by quantum repeater with respect to classical communication. The hybrid quantum repeater is investigated in \cite{Ladd2006,Van2006} using bright coherent light. A quantum repeater which uses photon pair sources in combination with memories has been proposed \cite{Simon2007}. Also, it is shown that the local generation of entangled pairs of atomic excitations, in combination with two-photon detections permits the implementation of a quantum repeater protocol \cite{Sangouard2008}. A quantum repeater model using  quantum-dot spins and  photons with spatial entanglement  in optical microcavities has been studied in \cite{Wang2012}. An experimental realization of entanglement concentration and a quantum repeater has been reported in \cite{Zhao2003}.\\
As two sticking points in the quantum repeater protocols one may refer to the entanglement swapping and quantum memories. The entanglement can be swapped in short parts of a long distance by performing interaction described by Tavis-Cummings \cite{Tavis1968} or Jaynes-Cummings \cite{Jaynes1963} models. After producing the entanglement, by Bell state measurement (BSM) \cite{weinfurter1994,Liao2011,Ghasemi2017} or performing interaction between separable parts \cite{Pakniat2017} the entanglement is swapped to non-entangled segments (for instance, see \cite{Zukowski1993,Ian2016,Lopez2017,Deng2017,Qin2018,Xie2016,Ghasemi2016,Nourmandipour2016,nourmandipour2015,xiao2008,ai2008}).\\
In this paper we consider a quantum repeater protocol to distribute entanglement between two distant atoms. To achieve the purpose, by inserting six atoms between the two far atoms we divide the distance to seven shorter parts (see figure \ref{fig:Fig1}). Then, by performing the interaction between atoms (2,3) and (6,7) in two two-mode cavities, the entanglement is produced in the two pairs (1,4) and (5,8), separately. 
At last, the atoms (1,8) can be entangled either by operating the BSM (performing interaction) between (on) atoms (4,5) in a single-mode cavity.
The degree of entanglement of the produced entangled states of atoms (1,8) is evaluated by concurrence \cite{Wootters1998}. In the continuation, we have analyzed the influence of atom-field coupling coefficient and detuning on the concurrence and success probability of the proposed model. Also, the effect of interaction time on the concurrence has been investigated.\\
This paper organizes as follows: In the next section we have introduced our  quantum repeater protocol model and presented the detailed calculations. The related numerical results are collected in Sec. 3. Finally, the paper is ended by conclusions in Sec. 4.

\section{2. Quantum repeater protocol}\label{sec.model}
 \begin{figure}[H]
   \centering
 \includegraphics[width=0.28\textwidth]{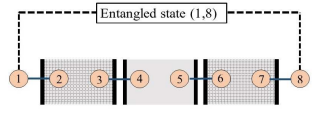}
   \caption{\label{fig:Fig1} {Quantum repeater protocol: there exist four atomic Bell states $(i,i+1)$, where $i=1,3,5,7$, the atomic pairs (2,3) and (6,7) interact in two-mode cavities, separately, and  two atoms (4,5) are contacted by operating Bell state measurement process or performing an interaction in a single-mode cavity.}}
  \end{figure}
As we demonstrated in the Introduction of the paper,  in the model shown in figure \ref{fig:Fig1} , our aim is the generation of entanglement between two separable distant atoms 1, 8. 
At first, we consider the interaction between atoms (2,3), where the initial state for atoms (1-4) reads as $\ket{\Psi}_{1,2}\otimes\ket{\Psi}_{3,4}$  with,
   \begin{eqnarray}\label{initialstate}
 \ket{\Psi}_{i,i+1}&=&\frac{1}{\sqrt{2}}(\ket{eg}-\ket{ge})_{i,i+1}, \qquad   i=1,3.
  \end{eqnarray}
  Two two-level atoms (2,3) with excited (ground) state $\ket{e}$ ($\ket{g}$) in a two-mode cavity field are interacted by the Tavis-Cummings Hamiltonian introduced by,
 \begin{equation}
 \hat{H}=\hat{H}_0+\hat{H}_1,
 \end{equation}
   where the free and interacting parts are respectively defined as $(\hbar=1)$ \cite{Obada2008}:
\begin{eqnarray}\label{hamiltonian}
\hat{H}_0&=&\omega\hat{a}^{\dagger}\hat{a}+\omega'\hat{b}^{\dagger}\hat{b}+\sum_{i=2,3} \frac{\omega_i}{2}\hat{\sigma}_{iz},  \\ \nonumber
\hat{H}_1&=&\sum_{i=2,3} g_i\left( \hat{a}\hat{b}\hat{\sigma}_i^++\hat{a}^{\dagger}\hat{b}^{\dagger}\hat{\sigma}_i^-\right),
\end{eqnarray}
where in fact we have considered the two-photon transition process in (\ref{hamiltonian}). It should be mentioned that the two-photon process \cite{Bashkirov2008,Zhou2002} is not far from experimental realization (as a few efforts in this relation see Refs. \cite{Deppe2008,Isenhower2010,Mohapatra2007}). In above relations, the creation (annihilation) operators of the two modes have been shown by $\hat{a}^{\dagger}$, $\hat{b}^{\dagger}$ ($\hat{a}$, $\hat{b}$) and $\hat{\sigma}_i^+=\ket{e}_i\bra{g}(\hat{\sigma}_i^-=\hat{\sigma}_i^{+\dagger})$, $\hat{\sigma}_{iz}$ are raising (lowering) and population inversion operators of the $i$th atom, respectively. The $i$th atom with frequency $\omega_i$ (in this case, $i=2,3$) interacts with two-mode field with frequencies $\omega$, $\omega'$. The corresponding Hamiltonian in the interaction picture reads as,
\begin{equation}
\hat{H}_{int}=\sum_{i=2,3}g_i\left( \hat{a}\hat{b}\hat{\sigma_i}^+ e^{i\Delta_i t}+H.C.\right),
\end{equation}
  which may be obtained from $\hat{H}_{int}=e^{i\hat{H}_0t}\hat{H}_1e^{-i\hat{H}_0t}$ by Baker-Hausdorff formula, where the detuning is defined as $\Delta_i=\omega_i-(\omega+\omega')$. We assume that the cavity is initially in vacuum state and also $\Delta_i\gg g_i$. So, following the path of Refs. \cite{James2007,Zheng2000} the effective Hamiltonian is obtained as follows:
  \small
\begin{eqnarray}\label{effectivehamiltonian}
\hat{H}_{eff}&=& g_{23} \left(\hat{\sigma}_2^+\hat{\sigma}_3^- e^{i(\Delta_2-\Delta_3)t}+H.C.\right)+\sum_{i=2,3} \lambda_i \hat{\sigma}_i^+\hat{\sigma}_i^- ,\\
\lambda_i&=&\frac{g^2_i}{\Delta_i}, \qquad \frac{1}{\bar{\omega}_{23}}=\frac{1}{2}(\frac{1}{\Delta_2}+\frac{1}{\Delta_3}), \qquad g_{23}=\frac{g_2 g_3}{\bar{\omega}_{23}}.
\end{eqnarray}
 Now, we calculate the time evolution operator with the help of effective Hamiltonian (\ref{effectivehamiltonian}) and using the relation $i \hat{\dot{U}}(t)=\hat{H}_{eff}\hat{U}(t)$ in the basis spanned by $\left\lbrace \ket{ee}_{2,3}, \ket{eg}_{2,3}, \ket{ge}_{2,3}, \ket{gg}_{2,3} \right\rbrace $,
\small
 \begin{equation}
    \hat{U}_{j,j+1}(t)= \left( \begin{array}{cccc}
     e^{-i(\lambda_2+\lambda_3) t} & 0 & 0 & 0 \\
     0 & e^{-i\lambda_2 t} e^{i \delta_{23} t/2}\left( \cos(\frac{ft}{2})-i\frac{\delta_{23}}{f} \sin(\frac{ft}{2})\right)  & e^{-i\lambda_2 t} e^{i \delta_{23} t/2}\left(-2i\frac{g_{23}}{f} \sin(\frac{ft}{2}) \right)  & 0  \\
     0 & e^{-i\lambda_3 t} e^{-i \delta_{23} t/2}\left(-2i\frac{g_{23}}{f} \sin(\frac{ft}{2}) \right) & e^{-i\lambda_3 t} e^{-i \delta_{23} t/2}\left( \cos(\frac{ft}{2})+i\frac{\delta_{23}}{f} \sin(\frac{ft}{2})\right) & 0  \\
     0 & 0 & 0 & 1  \\ \end{array} \right)_{j,j+1},
  \label{eq:unitarymatrix1}
  \end{equation}
 $$\mbox{\textit{see eq.~\eqref{eq:unitarymatrix1}}}$$
  where $j=2,6$ and
  \begin{eqnarray}
  \delta_{23}=\Delta_2-\Delta_3+(\lambda_2-\lambda_3),  \qquad    f=\sqrt{\delta_{23}^2+4g^2_{23}}.
  \end{eqnarray}
 Then, using the initial state $\ket{\Psi}_{1,2}\otimes\ket{\Psi}_{3,4}$ (Eq. (\ref{initialstate})) and (\ref{eq:unitarymatrix1}), the state of atoms (1,2,3,4) after the interaction at time $t$ can be deduced as below:
 \small
\begin{eqnarray}\label{state1-4}
 \ket{\Psi(t)}_{1-4}&=&\hat{U}_{2,3}(t)\ket{\Psi}_{1,2}\otimes\ket{\Psi}_{3,4}\\ \nonumber
 &=&\frac{1}{2}\left( U_{23}(t)\ket{eegg}+U_{33}(t)\ket{egeg}-\ket{egge}\right. \\ \nonumber
 &-&\left. U_{11}(t)\ket{geeg}+U_{22}(t)\ket{gege}+U_{32}(t)\ket{ggee}\right) _{1-4},
\end{eqnarray}
where $U_{ij}(t)$ is the matrix element of (\ref{eq:unitarymatrix1}) locating in the row $i$ and column $j$. If the atoms (2,3) be in $\ket{ge}_{2,3}$ or $\ket{eg}_{2,3}$, the state of atomic pair (1,4) is achieved respectively as follows,
\small
\begin{eqnarray}\label{state14}
   \ket{\Psi(t)}_{1,4}&=&\frac{1}{\sqrt{\left| U_{33}(t)\right|^2+\left| U_{32}(t)\right|^2} }(U_{33}(t)\ket{eg}\\ \nonumber
         &+& U_{32}(t)\ket{ge})_{1,4},\nonumber \\
   \ket{\Psi'(t)}_{1,4}&=&\frac{1}{\sqrt{\left| U_{23}(t)\right|^2+\left| U_{22}(t)\right|^2} }(U_{23}(t)\ket{eg}\\ \nonumber
         &+& U_{22}(t)\ket{ge})_{1,4}.
\end{eqnarray}
Similarly, atoms (5,6,7,8) are evolved with the effective Hamiltonian (\ref{effectivehamiltonian}) as above procedure, however,
with the time evolution operator $\hat{U}_{6,7}(t)$ in (\ref{eq:unitarymatrix1}), by which the entangled states of atoms (5,8) can be obtained as follows,
\small
\begin{eqnarray}\label{state58}
   \ket{\Psi(t)}_{5,8}&=&\frac{1}{\sqrt{\left| U_{33}(t)\right|^2 + \left| U_{32}(t)\right|^2} }(U_{33}(t)\ket{eg}\\ \nonumber
      &+&  U_{32}(t)\ket{ge})_{5,8},\nonumber \\
   \ket{\Psi'(t)}_{5,8}&=&\frac{1}{\sqrt{\left| U_{23}(t)\right|^2+\left| U_{22}(t)\right|^2} }(U_{23}(t)\ket{eg}\\ \nonumber
         &+& U_{22}(t)\ket{ge})_{5,8},
\end{eqnarray}
 depending on measuring $\ket{ge}_{6,7}$ and $\ket{eg}_{6,7}$, respectively. Consequently, we have four possible non-entangled states $\ket{\Psi(t)}_{1,4}\otimes \ket{\Psi(t)}_{5,8}$, $ \ket{\Psi(t)}_{1,4}\otimes \ket{\Psi'(t)}_{5,8}$, $ \ket{\Psi'(t)}_{1,4}\otimes \ket{\Psi(t)}_{5,8}$ and $ \ket{\Psi'(t)}_{1,4}\otimes \ket{\Psi'(t)}_{5,8}$. Considering all these states we are going to introduce our quantum repeater protocol and compare the obtained results.\\
\subsection{The case $\ket{\Psi(t)}_{1,4}\otimes \ket{\Psi(t)}_{5,8}$}
In this stage, we operate the Bell state measurement performed with the state $\ket{B}_{4,5}=\frac{1}{\sqrt{2}}(\ket{ee}+\ket{gg})_{4,5}$ on state $\ket{\Psi(t)}_{1,4}\otimes \ket{\Psi(t)}_{5,8}$. As a result, the state of atoms (1,8) is converted to the Bell state,
\begin{equation}\label{bell}
\ket{B}_{1,8}=\frac{1}{\sqrt{2}}(\ket{ee}+\ket{gg})_{1,8},
\end{equation}
with the success probability
  \begin{equation}\label{sucb}
   S_{\ket{B}_{1,8}}(t)=\frac{\left| U_{33}(t) U_{32}(t) \right|^2 }{(\left|U_{33}(t) \right|^2+ \left|U_{32}(t) \right|^2)^2}.
   \end{equation}
 Now, we perform a BSM with Bell state $\ket{B'}_{4,5}=\frac{1}{\sqrt{2}}(\ket{eg}+\ket{ge})_{4,5}$ on state $ \ket{\Psi(t)}_{1,4}\otimes \ket{\Psi(t)}_{5,8}$, so the atoms (1,8) are converted to the following entangled state,
 \begin{equation}
 \ket{\gamma}^1_{1,8}=\frac{1}{\sqrt{ \left| U_{33}(t) \right|^4+ \left| U_{32}(t) \right|^4}}(U^2_{33}(t)\ket{eg}+U^2_{32}(t)\ket{ge})_{1,8},
 \end{equation}
   whose concurrence reads as
   \begin{equation}
   C_1(t)=\frac{2\left| U^{*2}_{33}(t) U^2_{32}(t) \right| }{\left|U_{33}(t) \right|^4+ \left|U_{32}(t) \right|^4}.
   \end{equation}
Next, we consider the initial state $\ket{\Psi(t)}_{1,4}\otimes \ket{\Psi(t)}_{5,8}$ and investigate the interaction between atoms (4,5) in a single-mode cavity using the effective Hamiltonian,
\begin{equation}\label{efham}
\hat{H}^\prime_{eff}=\lambda'\sum_{i=4,5}\hat{\sigma}_i^+\hat{\sigma}_i^-+ \lambda'\left(\hat{\sigma}_4^+\hat{\sigma}_5^-+\hat{\sigma}_4^-\hat{\sigma}_5^+\right),
\end{equation}
  \begin{eqnarray}
  \lambda'=\frac{g^2}{\delta},  \qquad  \delta=\omega_4-\omega=\omega_5-\omega.
  \end{eqnarray}
 Notice that the above Hamiltonian has been readily obtained by using the path of Refs. \cite{James2007,Zheng2000} and the following Hamiltonian
 \begin{equation}
\hat{H'}=\omega\hat{a}^{\dagger}\hat{a}+\sum_{i=4,5} \frac{\omega_i}{2}\hat{\sigma}_{iz}+\sum_{i=4,5} g \left( \hat{a}\hat{\sigma}_i^++\hat{a}^{\dagger}\hat{\sigma}_i^-\right).
 \end{equation}
By performing the interaction (\ref{efham}) between atoms (4,5) for the initial state $ \ket{\Psi(t)}_{1,4}\otimes \ket{\Psi(t)}_{5,8}$, the following entangled state can be achieved for atoms $(1,4,5,8)$,
\begin{equation}\label{intstate1458}
\begin{aligned}
  \ket{\phi}_{1,4,5,8}&=\frac{1}{\left|U_{32}(t) \right|^2+\left|U_{33}(t) \right|^2 } \\
        &\left\lbrace U^2_{33}(t)\left[\frac{1}{2}(e^{2i\lambda' t}e^{-2i\lambda' \tau}-1) \ket{eegg}\right. \right. \\
              &+\left. \left. \frac{1}{2}(e^{2i\lambda' t}e^{-2i\lambda' \tau}+1)\ket{egeg}\right]\right.    \\
         & +\left.  U^2_{32}(t) \left[\frac{1}{2}(e^{2i\lambda' t}e^{-2i\lambda' \tau}+1) \ket{gege}\right. \right. \\
               &+\left. \left. \frac{1}{2}(e^{2i\lambda' t}e^{-2i\lambda' \tau}-1)\ket{ggee} \right]\right. \\
        & +\left.  U_{32}(t) U_{33}(t) \left(\ket{egge}\right. \right. \\
        &+\left. \left. e^{-2i\lambda' (\tau-t)} \ket{geeg} \right) \right\rbrace _{1,4,5,8}.
\end{aligned}
\end{equation}?
By making a measurement on the state (\ref{intstate1458}), one obtains $\ket{eg}_{4,5}$ and $\ket{ge}_{4,5}$, the result of which arrives us respectively at
 \begin{eqnarray}\label{intstate18}
    \ket{\phi}^1_{1,8}&= &\frac{1}{\sqrt{\left| a_1(t,\tau)\right|^2+\left| b_1(t,\tau)\right|^2 }}\left(a_1(t,\tau)\ket{eg}\right. \\ \nonumber
          &+& \left. b_1(t,\tau)\ket{ge} \right) _{1,8},\\ \nonumber
 a_1(t,\tau)&=&U^2_{33}(t)(e^{2i\lambda' t}e^{-2i\lambda' \tau}-1),\\ \nonumber
 b_1(t,\tau)&=&U^2_{32}(t)(e^{2i\lambda' t}e^{-2i\lambda' \tau}+1),
  \end{eqnarray}
with the concurrence
\begin{equation}
C'_1(t,\tau)=\frac{2\left|a^*_1(t,\tau) b_1(t,\tau) \right| }{\left| a_1(t,\tau)\right|^2+\left| b_1(t,\tau)\right|^2 },
\end{equation}
and
   \begin{eqnarray}\label{intstatee18}
      \ket{\phi'}^1_{1,8}&= &\frac{1}{\sqrt{\left| a'_1(t,\tau)\right|^2+\left| b'_1(t,\tau)\right|^2 }}\left(a'_1(t,\tau)\ket{eg}\right. \\ \nonumber
            &+&\left.  b'_1(t,\tau)\ket{ge} \right) _{1,8},\\ \nonumber
   a'_1(t,\tau)&=&U^2_{33}(t)(e^{2i\lambda' t}e^{-2i\lambda' \tau}+1),\\ \nonumber
   b'_1(t,\tau)&=&U^2_{32}(t)(e^{2i\lambda' t}e^{-2i\lambda' \tau}-1),
    \end{eqnarray}
with the concurrence
\begin{equation}
 C^{\prime\prime}_1(t,\tau)= \frac{2\left|a'^*_1(t,\tau) b'_1(t,\tau) \right| }{\left| a'_1(t,\tau)\right|^2+\left| b'_1(t,\tau)\right|^2 }.
\end{equation}
    \subsection{The case $ \ket{\Psi(t)}_{1,4}\otimes \ket{\Psi'(t)}_{5,8}$}
 By the operation of BSM performed with the state $\ket{B'}_{4,5}=\frac{1}{\sqrt{2}}(\ket{eg}+\ket{ge})_{4,5}$ on state $ \ket{\Psi(t)}_{1,4}\otimes \ket{\Psi'(t)}_{5,8}$, the atoms (1,8) are converted to the Bell state
 \begin{equation}\label{belll}
\ket{B'}_{1,8}=\frac{1}{\sqrt{2}}(\ket{eg}+e^{-2i\varphi}\ket{ge})_{1,8},
 \end{equation}
 where $e^{i\varphi}=\pm\frac{ \cos(\frac{ft}{2})+i\frac{\delta_{23}}{f} \sin(\frac{ft}{2})}{\left|  \cos(\frac{ft}{2})+i\frac{\delta_{23}}{f} \sin(\frac{ft}{2}) \right| }$ with $\left| e^{i\varphi} \right| =1$ and the success probability reads as
  \begin{equation}\label{sucb'}
 S_{\ket{B'}_{1,8}}(t)=\frac{\left| U_{33}(t) U_{23}(t)\right| ^2+\left|  U_{32}(t) U_{22}(t) \right|^2 }{2 (\left| U_{33}(t)\right| ^2+\left|  U_{32}(t) \right|^2) (\left| U_{23}(t)\right| ^2+\left|  U_{22}(t) \right|^2)}.
  \end{equation}
 Notice that  $S_{\ket{B'}_{1,8}}(t)= S_{\ket{B}_{1,8}}(t)$. But, by measuring the Bell state $\ket{B}_{4,5}=\frac{1}{\sqrt{2}}(\ket{ee}+\ket{gg})_{4,5}$ on state $ \ket{\Psi(t)}_{1,4}\otimes \ket{\Psi'(t)}_{5,8}$, the atoms (1,8) are converted to the entangled state
 \begin{eqnarray}
\ket{\gamma}^2_{1,8}&=\frac{1}{\sqrt{ \left| U_{22}(t) U_{33}(t) \right|^2+ \left| U_{23}(t) U_{32}(t) \right|^2}}\\ \nonumber
      &( U_{22}(t) U_{33}(t)\ket{ee}+ U_{23}(t) U_{32}(t)\ket{gg})_{1,8},
 \end{eqnarray}
 with the concurrence
 \begin{equation}
C_2(t)=\frac{2\left| U^*_{22}(t) U^*_{33}(t) U_{23}(t) U_{32}(t) \right| }{ \left| U_{22}(t) U_{33}(t) \right|^2+ \left| U_{23}(t) U_{32}(t) \right|^2}.
 \end{equation}
It can be seen that $C_2(t)=C_1(t)$.\\
 Now, the entangled state for atoms $(1,4,5,8)$ can be achieved after performing the interaction according to the Hamiltonian (\ref{efham}) between atoms (4,5) in a single-mode cavity with initial state $ \ket{\Psi(t)}_{1,4}\otimes \ket{\Psi'(t)}_{5,8}$ which results in,
\begin{equation}\label{intstate1458e}
\begin{aligned}
   \ket{\phi'}_{1,4,5,8}&=\frac{1}{\sqrt{\left|U_{32}(t) \right|^2+\left|U_{33}(t) \right|^2}\sqrt{\left|U_{22}(t) \right|^2+\left|U_{23}(t) \right|^2} }\\
   & \left\lbrace U_{33}(t)U_{23}(t)\left[\frac{1}{2}(e^{2i\lambda' t}e^{-2i\lambda' \tau}-1) \ket{eegg}\right. \right. \\
   &+\left. \left.  \frac{1}{2}(e^{2i\lambda' t}e^{-2i\lambda' \tau}+1)\ket{egeg}\right]\right.  \\
   &+\left.  U_{32}(t) U_{22}(t)\left[\frac{1}{2}(e^{2i\lambda' t}e^{-2i\lambda' \tau}+1) \ket{gege}\right. \right.  \\
   &+\left. \left.  \frac{1}{2}(e^{2i\lambda' t}e^{-2i\lambda' \tau}-1)\ket{ggee} \right]\right.  \\
         &+\left.  U_{33}(t) U_{22}(t) \ket{egge}\right. \\
         &+\left. e^{-2i\lambda' (\tau-t)}U_{32}(t) U_{23}(t) \ket{geeg} \right\rbrace _{1,4,5,8}.
\end{aligned}
\end{equation}?
By applying a measurement on the state (\ref{intstate1458e}), one obtains $\ket{eg}_{4,5}$ and $\ket{ge}_{4,5}$, where the results for atoms (1,8) are respectively as
  \begin{eqnarray}\label{intstate182}
     \ket{\phi}^2_{1,8}&= &\frac{1}{\sqrt{\left| a_2(t,\tau)\right|^2+\left| b_2(t,\tau)\right|^2 }}\left(a_2(t,\tau)\ket{eg}\right. \\ \nonumber
     &+&\left.  b_2(t,\tau)\ket{ge} \right) _{1,8},\\ \nonumber
  a_2(t,\tau)&=&U_{23}(t) U_{33}(t)(e^{2i\lambda' t}e^{-2i\lambda' \tau}-1),\\ \nonumber
  b_2(t,\tau)&=&U_{22}(t) U_{32}(t) (e^{2i\lambda' t}e^{-2i\lambda' \tau}+1),
   \end{eqnarray}
 whose concurrence reads as
    \begin{eqnarray}\label{concu2}
      C'_2(t,\tau)&= &\frac{2\left|a^*_2(t,\tau) b_2(t,\tau) \right| }{\left| a_2(t,\tau)\right|^2+\left| b_2(t,\tau)\right|^2 },
     \end{eqnarray}
   and
      \begin{eqnarray}\label{intstatee182}
          \ket{\phi'}^2_{1,8}&= &\frac{1}{\sqrt{\left| a'_2(t,\tau)\right|^2+\left| b'_2(t,\tau)\right|^2 }}\left(a'_2(t,\tau)\ket{eg}\right. \\ \nonumber
          &+& \left. b'_2(t,\tau)\ket{ge} \right) _{1,8},\\ \nonumber
       a'_2(t,\tau)&=&U_{23}(t) U_{33}(t) (e^{2i\lambda' t}e^{-2i\lambda' \tau}+1),\\ \nonumber
       b'_2(t,\tau)&=&U_{22}(t) U_{32}(t) (e^{2i\lambda' t}e^{-2i\lambda' \tau}-1),
        \end{eqnarray}
       with the concurrence
       \begin{equation}\label{c}
    C^{\prime\prime}_2(t,\tau)= \frac{2\left|a'^*_2(t,\tau) b'_2(t,\tau) \right| }{\left| a'_2(t,\tau)\right|^2+\left| b'_2(t,\tau)\right|^2 }.
       \end{equation}
   It is easy to check that $C^{\prime\prime}_2(t,\tau)=C'_2(t,\tau)$.
 \subsection{The case $ \ket{\Psi'(t)}_{1,4}\otimes \ket{\Psi(t)}_{5,8}$}
   The above processes can be repeated for the initial state $ \ket{\Psi'(t)}_{1,4}\otimes \ket{\Psi(t)}_{5,8}$. With a BSM using the maximally entangled state $\ket{B'}_{4,5}=\frac{1}{\sqrt{2}}(\ket{eg}+\ket{ge})_{4,5}$ on state $ \ket{\Psi'(t)}_{1,4}\otimes \ket{\Psi(t)}_{5,8}$, the state of atoms (1,8) is converted to Bell state $\ket{B'}_{1,8}$ in Eq. (\ref{belll}) with the success probability (\ref{sucb'}). Meanwhile, measuring the Bell state $\ket{B}_{4,5}=\frac{1}{\sqrt{2}}(\ket{ee}+\ket{gg})_{4,5}$ on state $ \ket{\Psi'(t)}_{1,4}\otimes \ket{\Psi(t)}_{5,8}$, the entangled state
   \begin{eqnarray}\label{es18}
\ket{\gamma}^3_{1,8}&= \frac{1}{\sqrt{ \left| U_{32}(t) U_{23}(t) \right|^2+ \left| U_{22}(t) U_{33}(t) \right|^2}}\\ \nonumber
&( U_{32}(t) U_{23}(t)\ket{ee}+ U_{22}(t) U_{33}(t)\ket{gg})_{1,8},
   \end{eqnarray}
 is obtained, where its concurrence is calculated as below:
 \begin{equation}
C_3(t)=\frac{2\left| U^*_{32}(t) U^*_{23}(t) U_{22}(t) U_{33}(t) \right| }{\left|U_{32}(t) U_{23}(t) \right|^2+ \left|U_{22}(t) U_{33}(t)\right|^2}.
 \end{equation}
   Notice that $C_3(t)=C_2(t)=C_1(t)$.\\
    On the other hand, the entangled state for atoms $(1,4,5,8)$ after performing the interaction according to the Hamiltonian (\ref{efham}) with initial state $ \ket{\Psi'(t)}_{1,4}\otimes \ket{\Psi(t)}_{5,8}$ can be obtained as
  \begin{equation}\label{intstate1458ee}
  \begin{aligned}
     \ket{\phi^{\prime\prime}}_{1,4,5,8}&=\frac{1}{\sqrt{\left|U_{32}(t) \right|^2+\left|U_{33}(t) \right|^2}\sqrt{\left|U_{22}(t) \right|^2+\left|U_{23}(t) \right|^2} } \\
     &\left\lbrace U_{33}(t)U_{23}(t)\left[\frac{1}{2}(e^{2i\lambda' t}e^{-2i\lambda' \tau}-1) \ket{eegg}\right. \right. \\
     &+\left. \left. \frac{1}{2}(e^{2i\lambda' t}e^{-2i\lambda' \tau}+1)\ket{egeg}\right]\right. \\
     &+ \left.  U_{32}(t) U_{22}(t)\left[\frac{1}{2}(e^{2i\lambda' t}e^{-2i\lambda' \tau}+1) \ket{gege}\right. \right. \\
     &+\left. \left. \frac{1}{2}(e^{2i\lambda' t}e^{-2i\lambda' \tau}-1)\ket{ggee} \right]\right.  \\
           &+\left.  U_{23}(t) U_{32}(t) \ket{egge}\right. \\
           &+\left. e^{-2i\lambda' (\tau-t)}U_{22}(t) U_{33}(t) \ket{geeg} \right\rbrace _{1,4,5,8}.
  \end{aligned}
  \end{equation}?
  If, the output of measurement on the state (\ref{intstate1458ee}) reads as $\ket{eg}_{4,5}$ and $\ket{ge}_{4,5}$, then the entangled states for atoms $(1,8)$ are achieved as $\ket{\phi}^3_{1,8}$ and $ \ket{\phi'}^3_{1,8}$ where $\ket{\phi}^3_{1,8}= \ket{\phi}^2_{1,8}$ and  $ \ket{\phi'}^3_{1,8}= \ket{\phi'}^2_{1,8}$ in Eqs. (\ref{intstate182}) and (\ref{intstatee182}) with the concurrences as $C'_3(t)=C'_2(t)$ and $C^{\prime\prime}_3(t)=C^{\prime\prime}_2(t)$ in Eqs. (\ref{concu2}) and (\ref{c}), respectively (notice that $C'_3(t)=C^{\prime\prime}_3(t)=C'_2(t)=C^{\prime\prime}_2(t)$).\\
   \subsection{The case $ \ket{\Psi'(t)}_{1,4}\otimes \ket{\Psi'(t)}_{5,8}$}
   For this case, after a BSM with $  \ket{B'}_{4,5}=\frac{1}{\sqrt{2}}(\ket{eg}+\ket{ge})_{4,5}$ on state $ \ket{\Psi'(t)}_{1,4}\otimes \ket{\Psi'(t)}_{5,8}$, the entangled state for atoms (1,8) is achieved as
   \begin{equation}
 \ket{\gamma}^4_{1,8}= \frac{1}{\sqrt{ \left| U_{23}(t) \right|^4+ \left| U_{22}(t)  \right|^4}}( U^2_{23}(t)\ket{eg}+ U^2_{22}(t)\ket{ge})_{1,8},
   \end{equation}
   with the concurrence
   \begin{equation}
  C_4(t)=\frac{2\left| U^{*2}_{23}(t) U^2_{22}(t) \right| }{\left|U_{23}(t) \right|^4+ \left|U_{22}(t) \right|^4}.
   \end{equation}
   Notice that $C_4(t)=C_3(t)=C_2(t)=C_1(t)$.\\
    Also, using of BSM performed with $\ket{B}_{4,5}=\frac{1}{\sqrt{2}}(\ket{ee}+\ket{gg})_{4,5}$ on state $ \ket{\Psi'(t)}_{1,4}\otimes \ket{\Psi'(t)}_{5,8}$, the Bell state (\ref{bell}) is obtained for atoms (1,8) with the success probability
     \begin{equation}
       S'_{\ket{B}_{1,8}}(t)=\frac{\left| U_{23}(t) U_{22}(t) \right|^2 }{(\left|U_{23}(t) \right|^2+ \left|U_{22}(t) \right|^2)^2}.
       \end{equation}
  Notice that $S'_{\ket{B}_{1,8}}(t)=S_{\ket{B}_{1,8}}(t)=S_{\ket{B'}_{1,8}}(t)$.\\
Moreover, by performing the interaction according to the Hamiltonian (\ref{efham}) between atoms (4,5) with initial state $ \ket{\Psi'(t)}_{1,4}\otimes \ket{\Psi'(t)}_{5,8}$, the following entangled state of atoms (1,4,5,8) is obtained as below:
  \begin{equation}\label{intstate14584}
  \begin{aligned}
     \ket{\phi^{\prime\prime\prime}}_{1,4,5,8}&=\frac{1}{\left|U_{23}(t) \right|^2+\left|U_{22}(t) \right|^2 }\\
     &\left\lbrace U^2_{23}(t)\left[\frac{1}{2}(e^{2i\lambda' t}e^{-2i\lambda' \tau}-1) \ket{eegg}\right. \right. \\
     &+\left. \left. \frac{1}{2}(e^{2i\lambda' t}e^{-2i\lambda' \tau}+1)\ket{egeg}\right]\right.  \\
     &+\left.  U^2_{22}(t)\left[\frac{1}{2}(e^{2i\lambda' t}e^{-2i\lambda' \tau}+1) \ket{gege}\right. \right. \\
     &+\left. \left. \frac{1}{2}(e^{2i\lambda' t}e^{-2i\lambda' \tau}-1)\ket{ggee} \right]\right. \\
     &+ \left.  U_{22}(t) U_{23}(t)\left(\ket{egge}\right.\right.  \\
     &+\left.\left.  e^{-2i\lambda' (\tau-t)} \ket{geeg} \right) \right\rbrace _{1,4,5,8}.
 \end{aligned}
 \end{equation}?
  Now, via measuring $\ket{eg}_{4,5}$ and $\ket{ge}_{4,5}$ on the obtained state in (\ref{intstate14584}), the atoms (1,8) are respectively converted to the states,
   \begin{eqnarray}\label{intstate184}
      \ket{\phi}^4_{1,8}&= &\frac{1}{\sqrt{\left| a_4(t,\tau)\right|^2+\left| b_4(t,\tau)\right|^2 }}\left(a_4(t,\tau)\ket{eg}\right. \\ \nonumber
      &+& \left. b_4(t,\tau)\ket{ge} \right) _{1,8},\\ \nonumber
   a_4(t,\tau)&=& U^2_{23}(t)(e^{2i\lambda' t}e^{-2i\lambda' \tau}-1),\\ \nonumber
   b_4(t,\tau)&=&U^2_{22}(t)(e^{2i\lambda' t}e^{-2i\lambda' \tau}+1),
    \end{eqnarray}
  with the  concurrence
  \begin{equation}
C'_4(t,\tau)= \frac{2\left|a^*_4(t,\tau) b_4(t,\tau) \right| }{\left| a_4(t,\tau)\right|^2+\left| b_4(t,\tau)\right|^2 },
  \end{equation}
 and
   \begin{eqnarray}\label{intstatee184}
      \ket{\phi'}^4_{1,8}&= &\frac{1}{\sqrt{\left| a'_4(t,\tau)\right|^2+\left| b'_4(t,\tau)\right|^2 }}\left(a'_4(t,\tau)\ket{eg}\right. \\ \nonumber
      &+& \left. b'_4(t,\tau)\ket{ge} \right) _{1,8},\\ \nonumber
   a'_4(t,\tau)&=&U^2_{23}(t) (e^{2i\lambda' t}e^{-2i\lambda' \tau}+1),\\ \nonumber
   b'_4(t,\tau)&=&U^2_{22}(t) (e^{2i\lambda' t}e^{-2i\lambda' \tau}-1),
    \end{eqnarray}
  with the concurrence
  \begin{equation}
C^{\prime\prime}_4(t,\tau)= \frac{2\left|a'^*_4(t,\tau) b'_4(t,\tau) \right| }{\left| a'_4(t,\tau)\right|^2+\left| b'_4(t,\tau)\right|^2 }.
  \end{equation}
  Notice that $C'_4(t,\tau)=C^{\prime\prime}_1(t,\tau)$ and $C^{\prime\prime}_4(t,\tau)=C'_1(t,\tau)$.
 \section{3. Results and discussion} \label{sec.results}
  \begin{figure}[h]
  \centering
  \subfigure[\label{fig.Fig2a} \ ]{\includegraphics[width=0.28\textwidth]{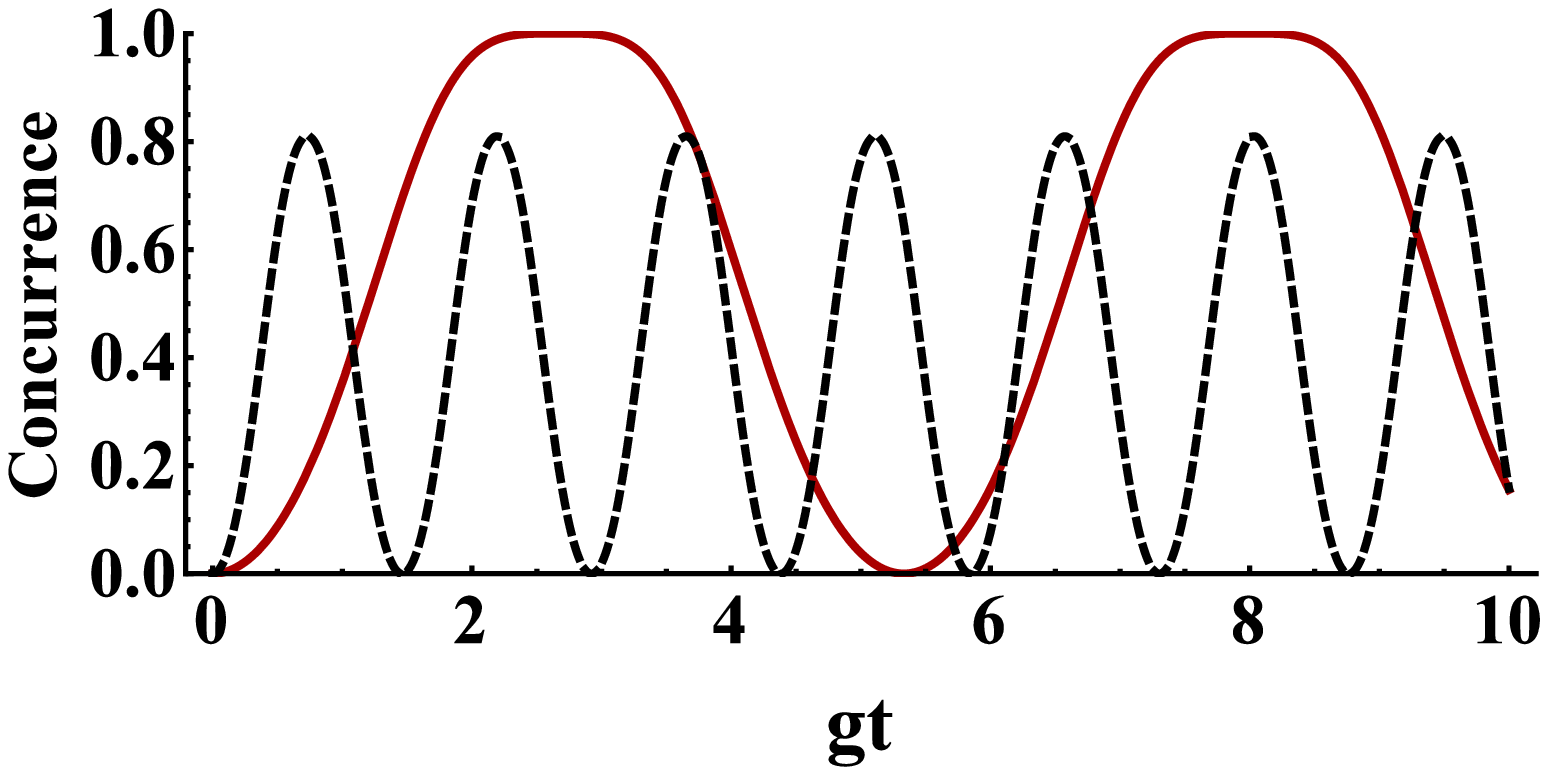}}
  \hspace{0.05\textwidth}
  \subfigure[\label{fig.Fig2b} \ ]{\includegraphics[width=0.28\textwidth]{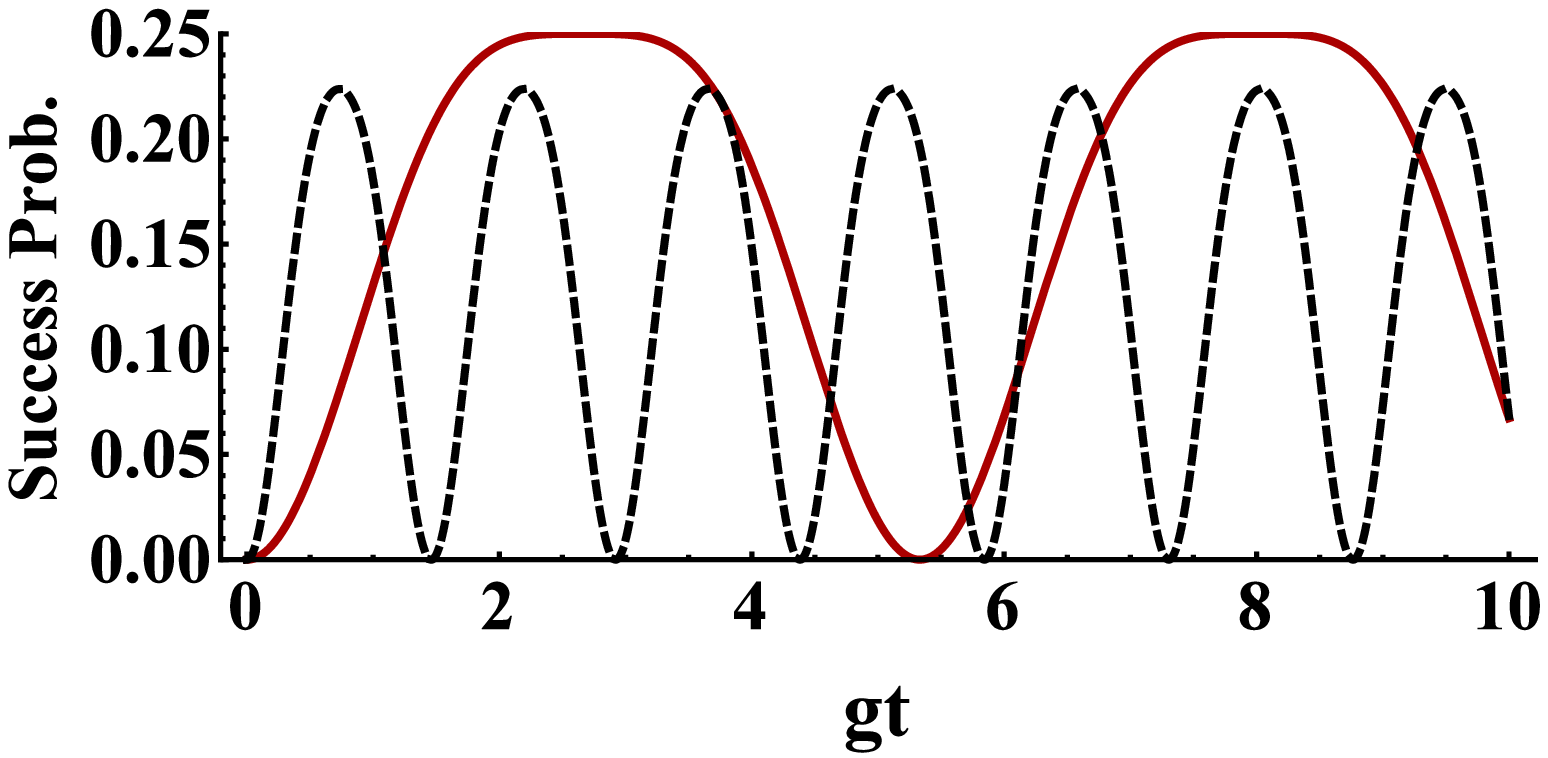}}

  \caption{\label{fig.csg} {\it The effect of coupling coefficient on the time evolution of}: (a) concurrence $C_1(t)=C_2(t)=C_3(t)=C_4(t)$ (b) success probability $S_{\ket{B}_{1,8}}(t)= S_{\ket{B'}_{1,8}}(t)=S'_{\ket{B}_{1,8}}(t)$, for symmetric condition $g=g_2=g_3$ (solid red line) and asymmetric condition $g_2=g$, $g_3=3g$ (dashed black line) with $\Delta_2=2g$ and $\Delta_3=3g$ (BSM method).}
  \end{figure}
   \begin{figure}[h]
   \centering
   \subfigure[\label{fig.Fig3a} \ ]{\includegraphics[width=0.28\textwidth]{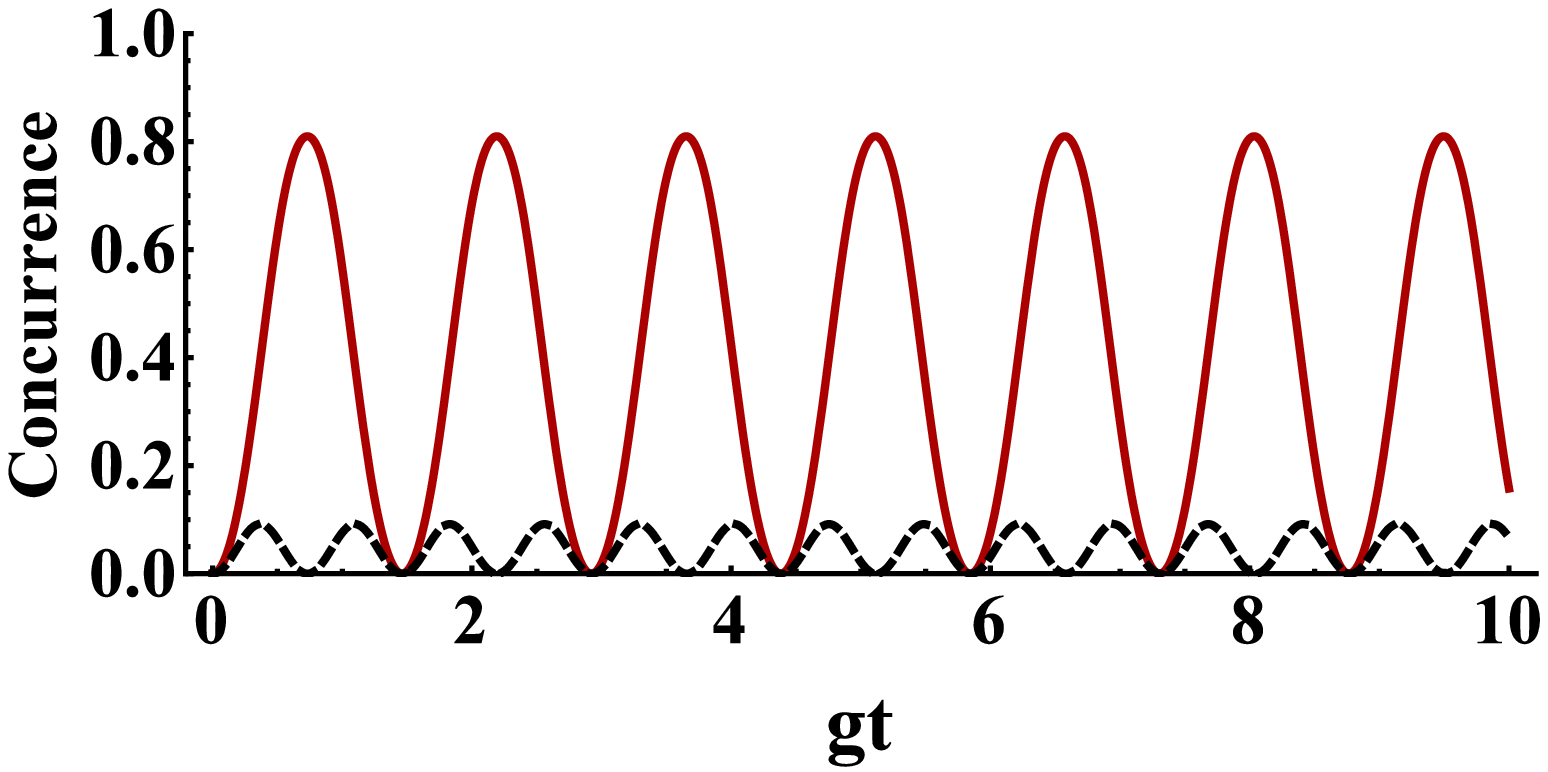}}
   \hspace{0.05\textwidth}
   \subfigure[\label{fig.Fig3b} \ ]{\includegraphics[width=0.28\textwidth]{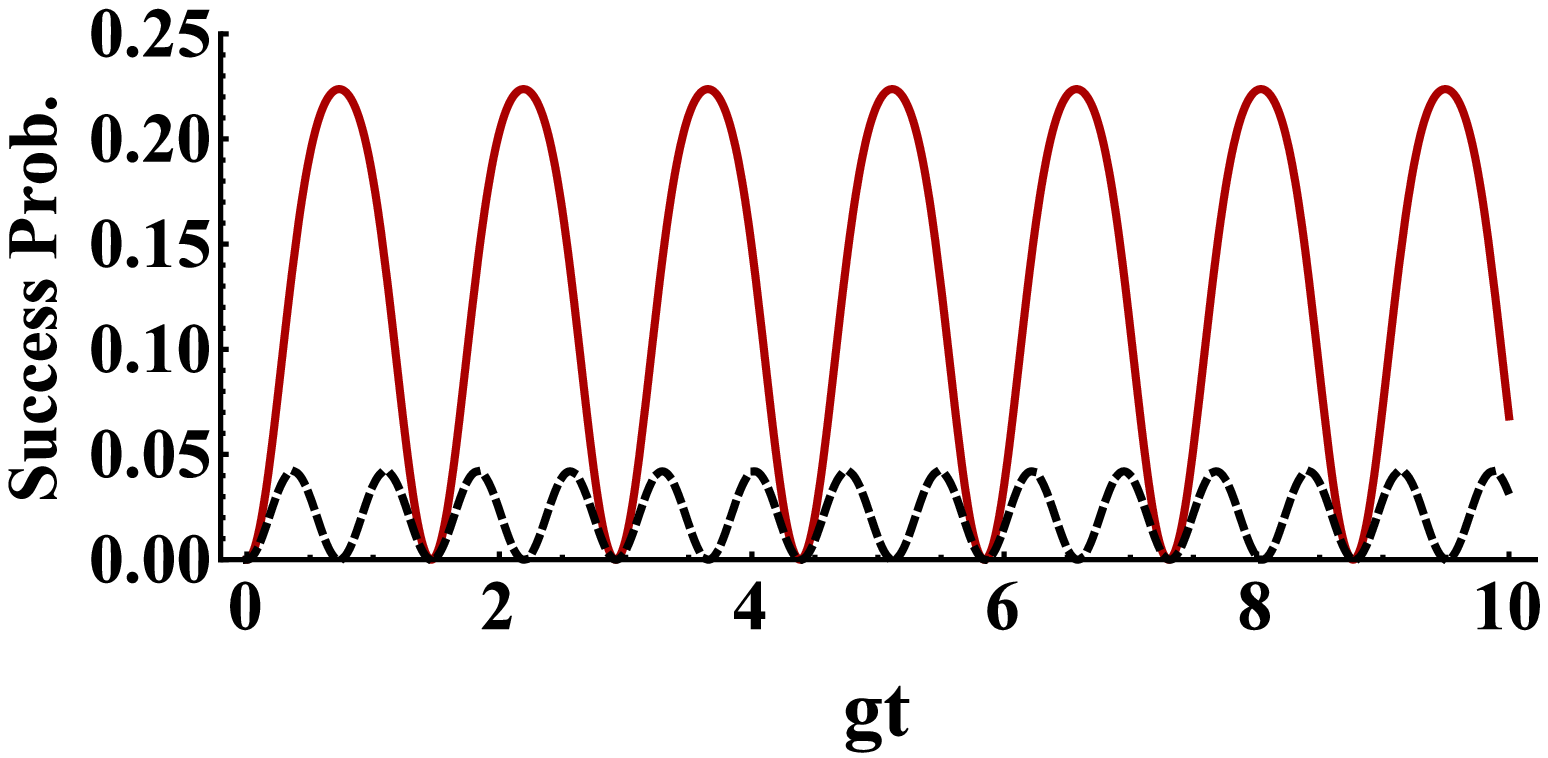}}

   \caption{\label{fig.csdet} {\it The effect of detuning on the time evolution of}: (a) concurrence $C_1(t)=C_2(t)=C_3(t)=C_4(t)$ (b) success probability $S_{\ket{B}_{1,8}}(t)= S_{\ket{B'}_{1,8}}(t)=S'_{\ket{B}_{1,8}}(t)$, for $\Delta_3=3g$ (solid red line) and $\Delta_3=10g$ (dashed black line) with $\Delta_2=2g$ for asymmetric condition $g_2=g$, $g_3=3g$ (BSM method).}
   \end{figure}

   \begin{figure}[h]
   \centering
   \subfigure[\label{fig.Fig4a} \ ]{\includegraphics[width=0.28\textwidth]{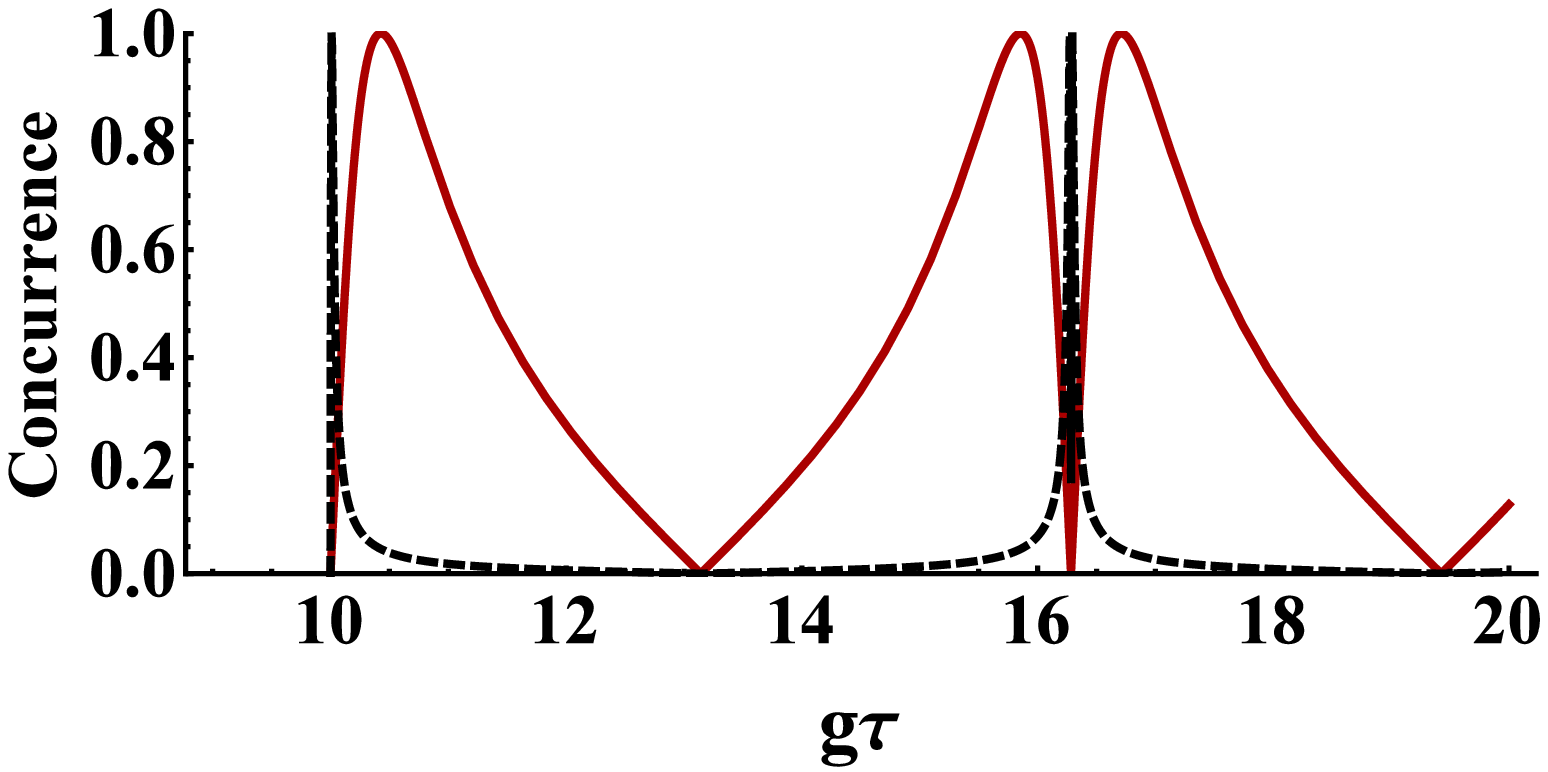}}
   \hspace{0.05\textwidth}
   \subfigure[\label{fig.Fig4b} \ ]{\includegraphics[width=0.28\textwidth]{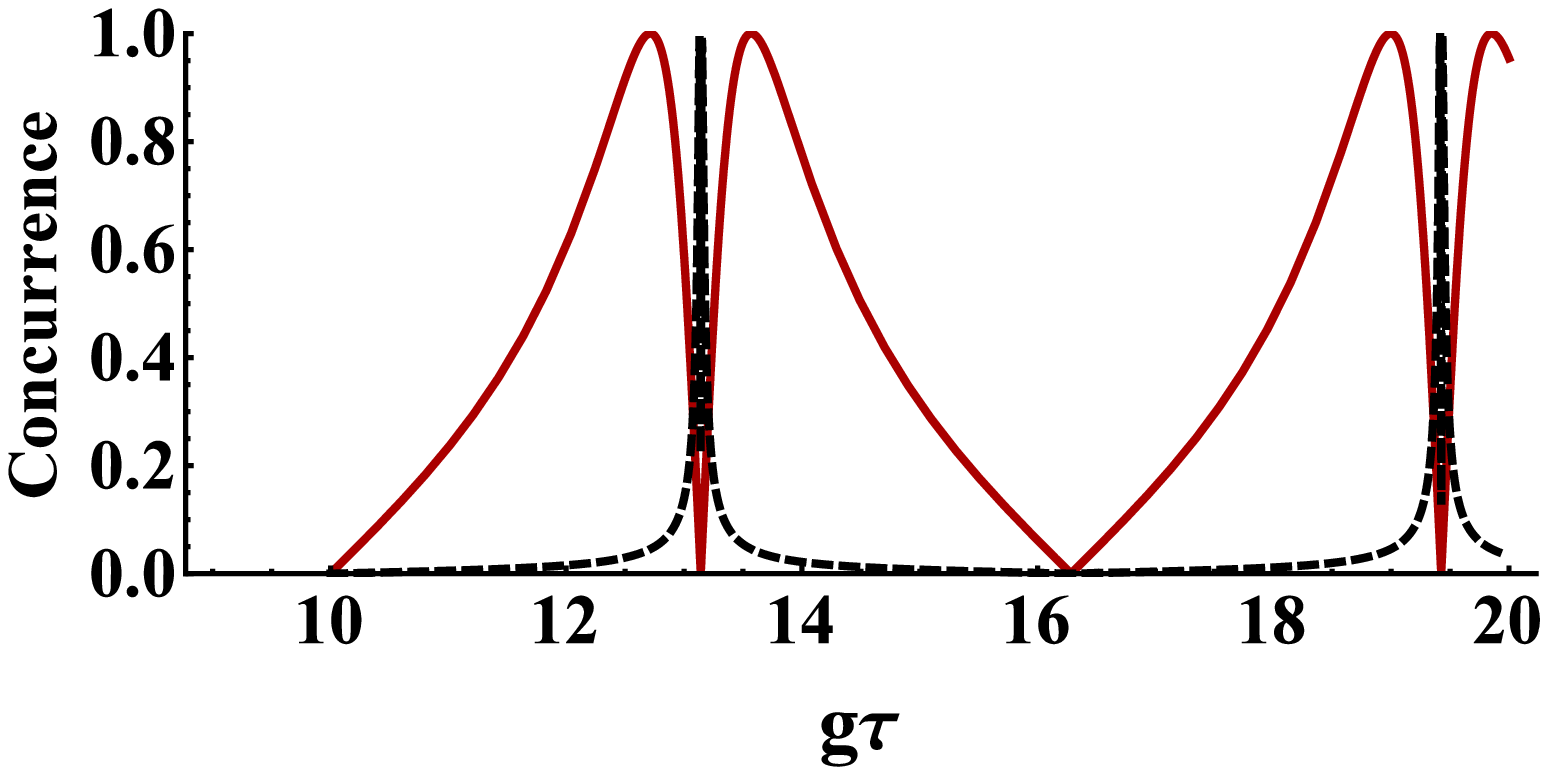}}
    \hspace{0.05\textwidth}
    \subfigure[\label{fig.Fig4c} \  ]{\includegraphics[width=0.28\textwidth]{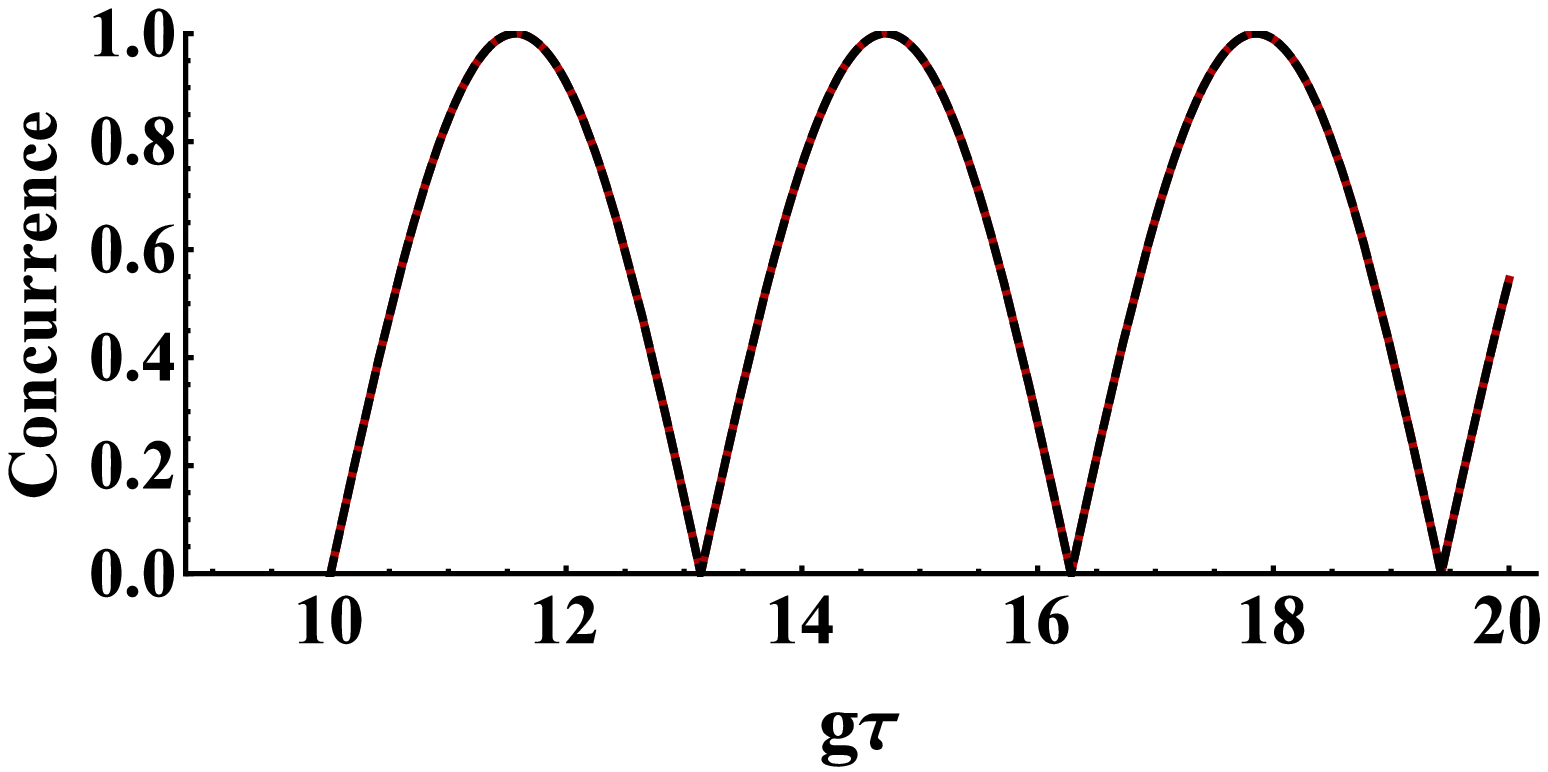}}

   \caption{\label{fig.cd} {\it The effect of detuning on the time evolution of concurrence}: (a) $C'_1(t,\tau)=C^{\prime\prime}_4(t,\tau)$ (b) $C'_4(t,\tau)=C^{\prime\prime}_1(t,\tau)$ (c) $C'_2(t,\tau)=C^{\prime\prime}_2(t,\tau)=C'_3(t,\tau)=C^{\prime\prime}_3(t,\tau)$, for  $\Delta_3=3g$ (solid red line) and  $\Delta_3=20 g$ (dashed black line) with $\delta=\Delta_2=2g$, $g_2=g$, $g_3=5g$ and $gt=10$ (QED method).}
   \end{figure}

   \begin{figure}[h]
   \centering
   \subfigure[\label{fig.Fig5a} \ ]{\includegraphics[width=0.28\textwidth]{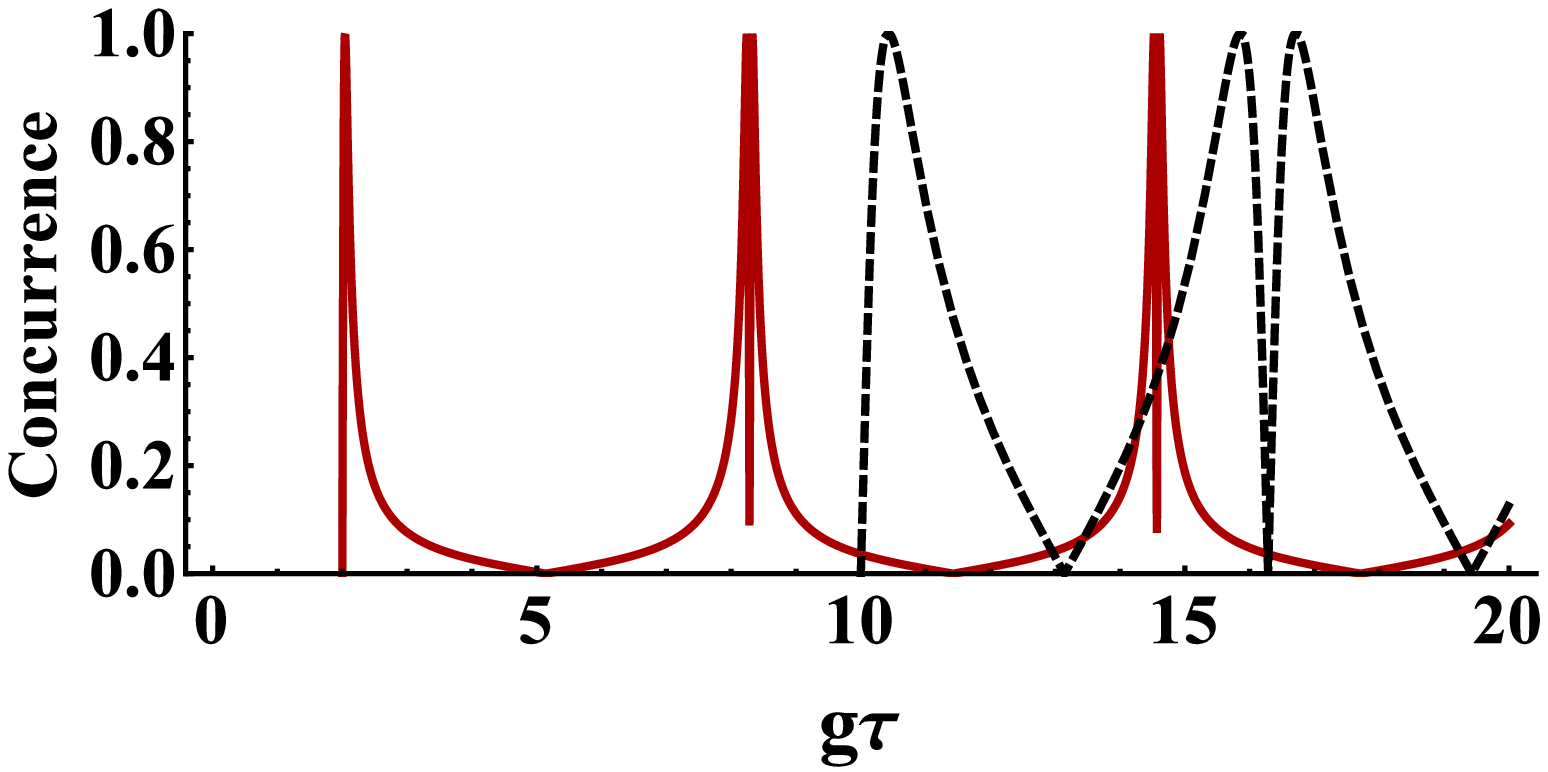}}
   \hspace{0.05\textwidth}
   \subfigure[\label{fig.Fig5b} \ ]{\includegraphics[width=0.28\textwidth]{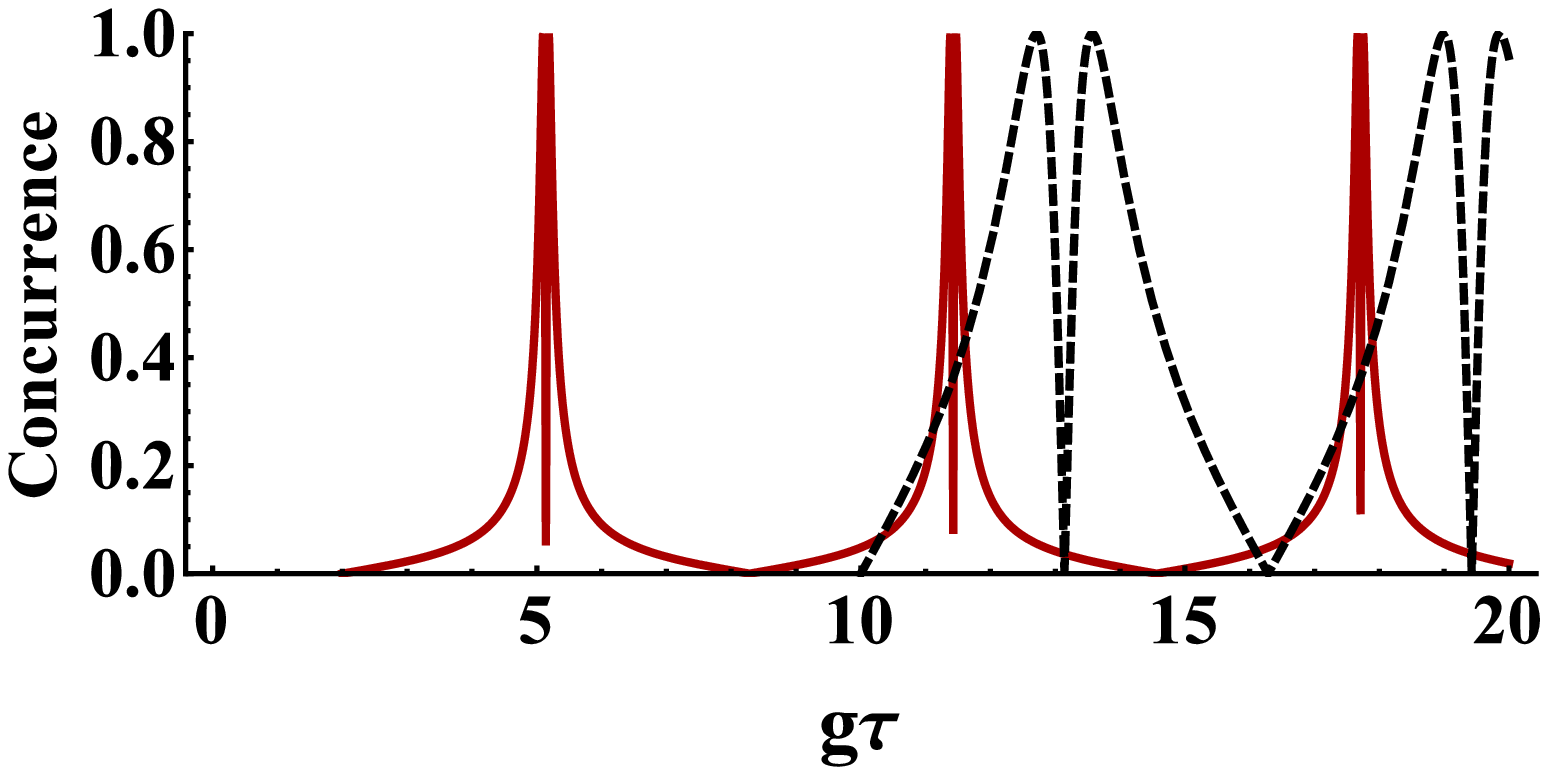}}
    \hspace{0.05\textwidth}
    \subfigure[\label{fig.Fig5c} \  ]{\includegraphics[width=0.28\textwidth]{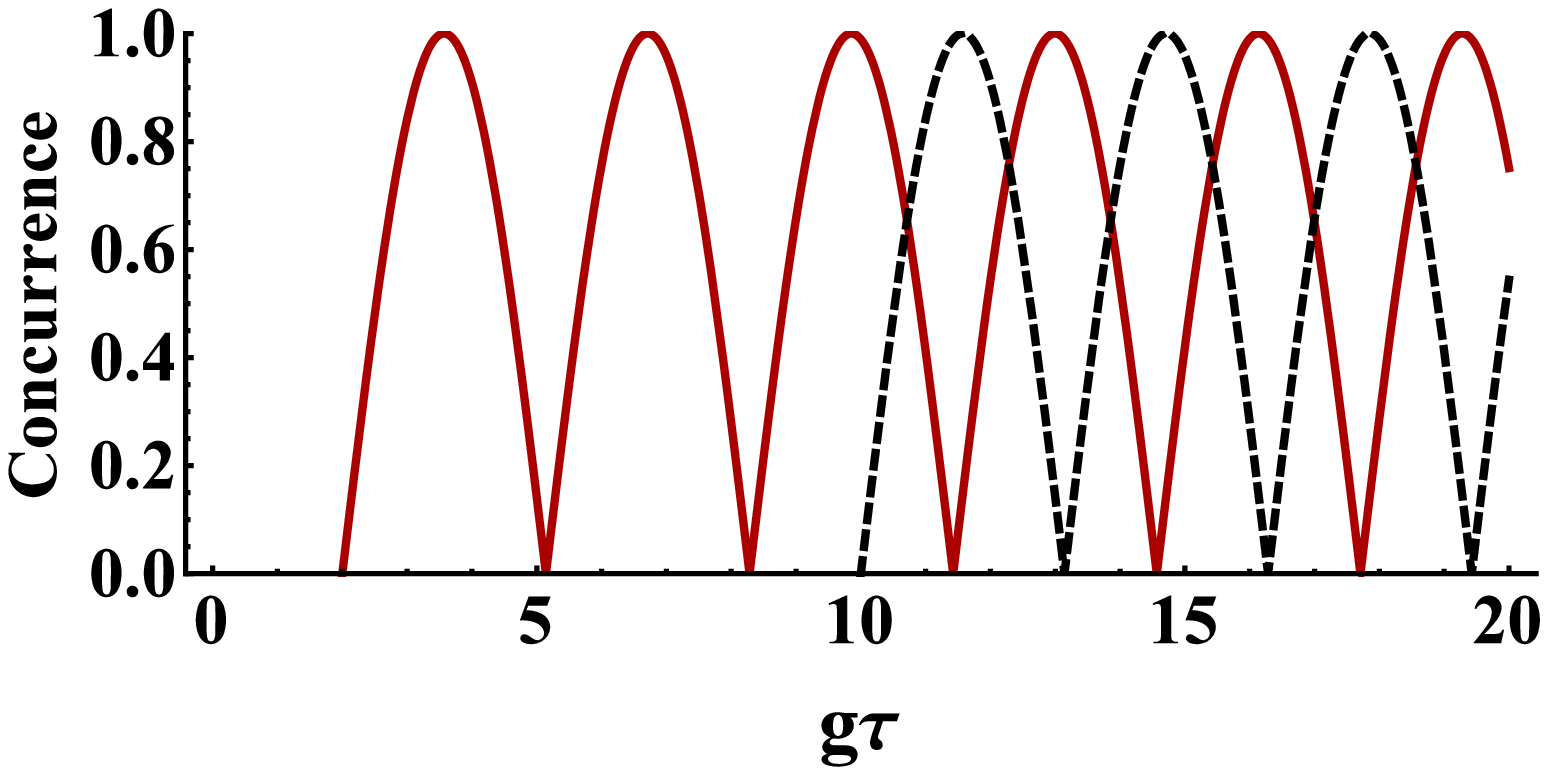}}

   \caption{\label{fig.cgt} {\it The effect of interaction time on the time evolution of concurrence}: (a) $C'_1(t,\tau)=C^{\prime\prime}_4(t,\tau)$ (b) $C'_4(t,\tau)=C^{\prime\prime}_1(t,\tau)$ (c) $C'_2(t,\tau)=C^{\prime\prime}_2(t,\tau)=C'_3(t,\tau)=C^{\prime\prime}_3(t,\tau)$, for $gt=2$ (solid red line) and  $gt=10$ (dashed black line) with $\delta=\Delta_2=2g$, $\Delta_3=3g$, $g_2=g$ and $g_3=5g$ (QED method).}
   \end{figure}

    \begin{figure}[h]
     \centering
     \subfigure[\label{fig.Fig6a} \ ]{\includegraphics[width=0.25\textwidth]{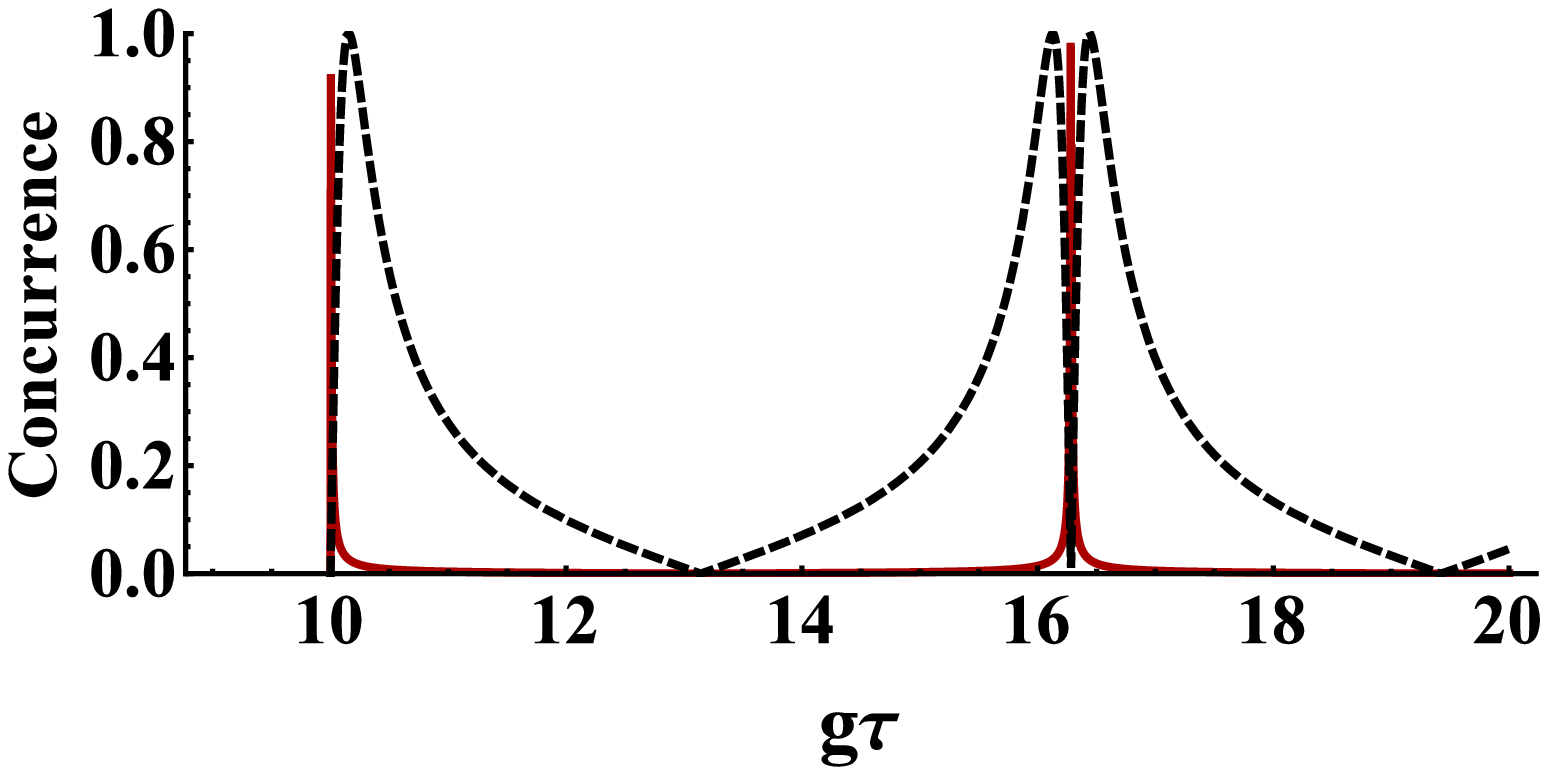}}
     \hspace{0.05\textwidth}
     \subfigure[\label{fig.Fig6b} \ ]{\includegraphics[width=0.25\textwidth]{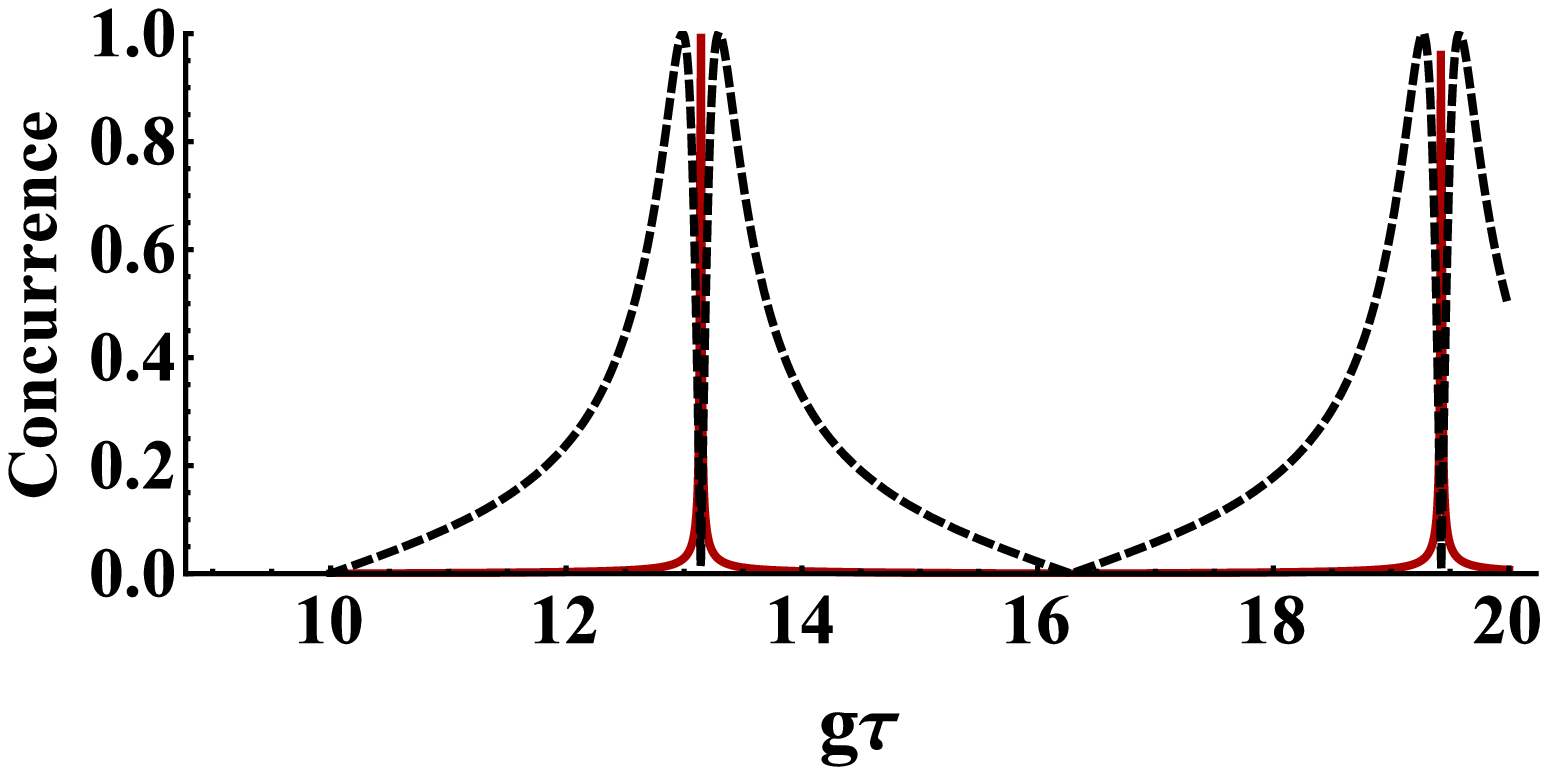}}
      \hspace{0.05\textwidth}
      \subfigure[\label{fig.Fig6c} \  ]{\includegraphics[width=0.25\textwidth]{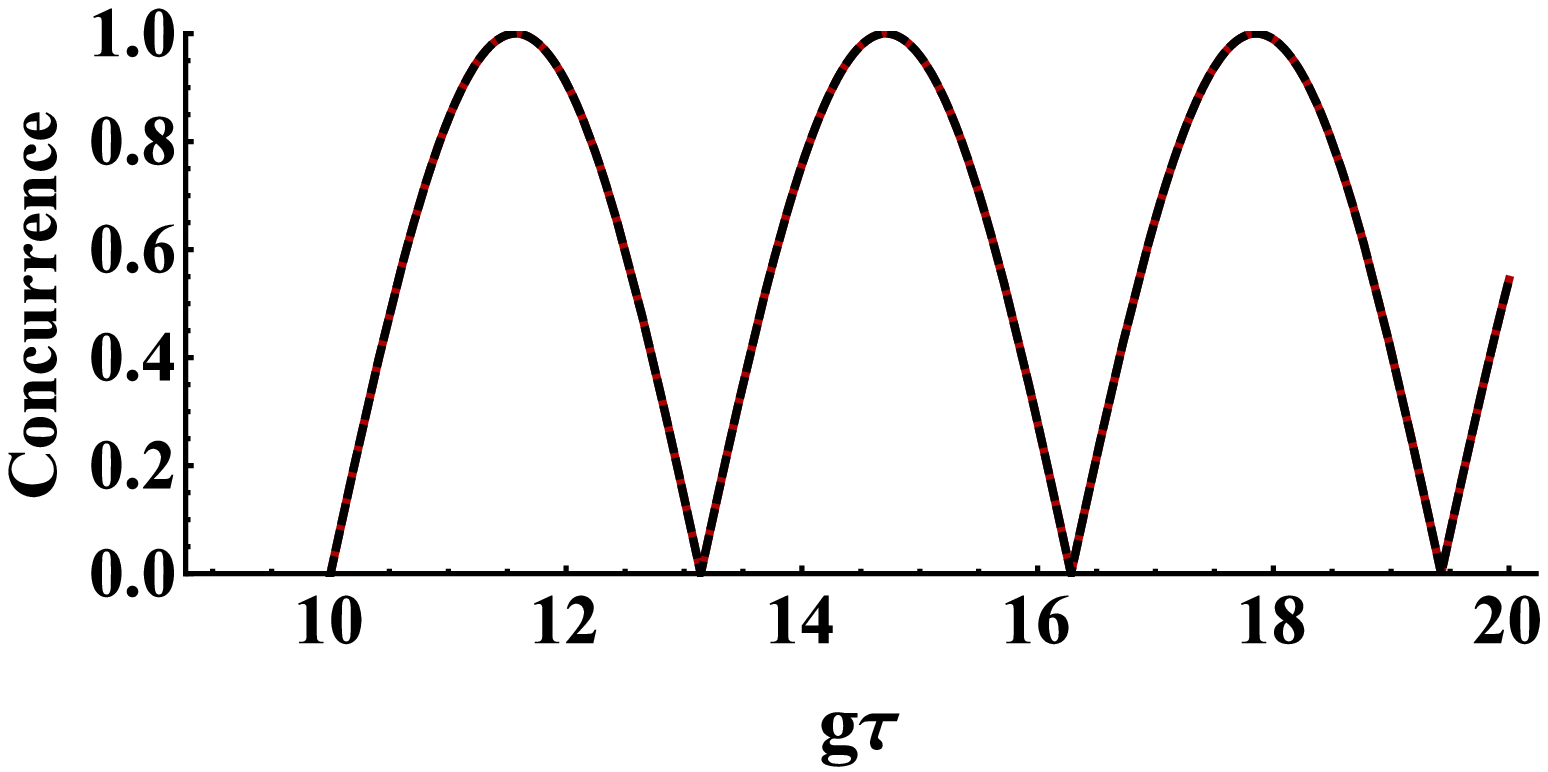}}

     \caption{\label{fig.cg} {\it The effect of coupling coefficient on the time evolution of concurrence}: (a) $C'_1(t,\tau)=C^{\prime\prime}_4(t,\tau)$ (b) $C'_4(t,\tau)=C^{\prime\prime}_1(t,\tau)$ (c) $C'_2(t,\tau)=C^{\prime\prime}_2(t,\tau)=C'_3(t,\tau)=C^{\prime\prime}_3(t,\tau)$, for $g_3=g$ (solid red line) and  $g_3=5g$ (dashed black line) with $\delta=\Delta_2=2g$, $\Delta_3=10g$, $g_2=g$ and $gt=10$ (QED method).}
     \end{figure}
In this section we present our numerical results and investigate about the degree of entanglement of the entangled states and success probabilities have been calculated in previous section. We have shown the time evolution of concurrence and success probability of entangled states for atoms $(1,8)$ produced by BSM method, where we have considered the effects of atom-field coupling coefficient and detuning on concurrence and success probability in figures \ref{fig.csg} and \ref{fig.csdet}, respectively. In figure \ref{fig.Fig2a} concurrence has been reached to its maximum value in some intervals of time, in symmetric condition, however, in asymmetric condition the maximum of concurrence and its time period have been decreased. The success probability in figure \ref{fig.Fig2b} has been reached to $0.25$ in some intervals of time for symmetric condition and in asymmetric condition this value and its time period have been decreased.\\
 In figure \ref{fig.csdet} we have considered the effect of detuning on concurrence and success probability. In figures \ref{fig.Fig3a} and \ref{fig.Fig3b} the maximum of concurrence  and success probability have been considerably decreased by increasing the detuning. Also, the time period of both quantities for $\Delta_3=3g$ is twice of the case $\Delta_3=10g$.\\
  The effect of detuning on the concurrence of entangled atoms $(1,8)$ produced by QED method has been considered in figure \ref{fig.cd}. In figures \ref{fig.Fig4a} and \ref{fig.Fig4b} the death of entanglement has been occurred in large time intervals by increasing the detuning in asymmetric condition and the entangled state of atoms $(1,8)$ has been converted to atomic Bell state for decreased detuning in more times. Varying detuning has no observable effect on concurrence in figure \ref{fig.Fig4c}; in addition, the death of entanglement has been occurred only for some instants of time.\\
    We have considered the effect of interaction time on the concurrence in figure \ref{fig.cgt}. The entanglement attenuation has been occurred in large time intervals for decreased interaction time in figures \ref{fig.Fig5a} and \ref{fig.Fig5b}. In figure \ref{fig.Fig5c} we can see similar evolution of concurrence for the two cases, but, maxima and minima of concurrence have been relocated.\\
     The effects of symmetric and asymmetric conditions have been considered in figure \ref{fig.cg}. The maxima of concurrence have been occurred in more times for asymmetric condition in figures \ref{fig.Fig6a} and \ref{fig.Fig6b}. Also, the death of entanglement has been occurred for some finite time intervals and then it suddenly revives in symmetric condition. In figure \ref{fig.Fig6c} it is shown that the evolution of concurrence is independent of coupling coefficient.
\section{4. Summary and conclusions} \label{sec.Conclusion}
In this paper we considered the quantum repeater protocol using mixed single- and two-mode Tavis-Cummings models. We investigated eight two-level atoms prepared in four entangled pairs (1,2), (3,4), (5,6) and (7,8). By performing the interaction between atoms (2,3) and (6,7) in two-mode cavities, the entangled atoms (1,4) and (5,8) were produced. Then, by using a BSM or performing the interaction between atoms (4,5) in a single-mode cavity (QED method), the atoms (1,8) were converted to an entangled state. The atomic states $(1,8)$ are converted to atomic Bell states by choosing suitable Bell states in BSM method. The effect of coupling coefficient and detuning have been investigated on the concurrence and success probability of entangled state of atoms (1,8) produced by BSM method. In asymmetric condition the maxima of concurrence and success probability and their time periods have been decreased by increasing the detuning in BSM method. Also, we considered the effects of detuning, interaction time and coupling coefficient on entanglement produced by QED method. We observed the destructive effects of increased detuning, \textit{i.e.}, the maxima of concurrence were occurred in less times and the death of entanglement was happened in large time interval when the detuning is increased. The effect of decreasing the interaction time $gt$ on the entanglement has been discussed, \textit{i.e.}, the entanglement attenuation was happened in large intervals of time by decreasing the interaction time. The death of entanglement in symmetric condition has been observed in large time interval for atoms (1,8) which are produced by QED method.\\
{\bf Acknowledgement}: The authors would like to thank the referees for their valuable comments which improved the paper.


\begin{thebibliography}{0}
\bibitem{Qin2017}
 Qin W., Wang X., Miranowicz A., Zhong Z. and Nori F.
 Phys. Rev. A 96 2017 012315.

\bibitem{Briegel1998}
 Briegel H-J, D{\"u}r W., Cirac J. I. and Zoller P.
 Phys. Rev. Lett.{81}{1998}{5932}.

\bibitem{Li2016}
 Li T., Yang G-J and Deng F-G
 Phys. Rev. A{93}{2016}{012302}.

 \bibitem{Rubenok2013}
  Rubenok A., Slater J. A., Chan P., Lucio-Martinez I. and Tittel W.
  Phys. Rev. Lett.{111}{2013}{130501}.

 \bibitem{muller1996}
 Muller A., Zbinden H. and Gisin N.
  Europhys. Lett.{33}{1996}{335}.


\bibitem{Su2018}
    Su Z., Guan J. and Li L.
    Phys. Rev. A{97}{2018}{012325}.

\bibitem{Ladd2006}
    Ladd T. D., van Loock P., Nemoto K., Munro W. J. and Yamamoto Y.
     New J. Phys.{8}{2006}{184}.

\bibitem{Van2006}
    Van Loock P., Ladd T. D., Sanaka K., Yamaguchi F., Nemoto K., Munro W. J. and Yamamoto Y.
     Phys. Rev. Lett.{96}{2006}{240501}.

\bibitem{Simon2007}
   Simon C., De Riedmatten H., Afzelius M., Sangouard N., Zbinden H. and Gisin N.
    Phys. Rev. Lett.{98}{2007}{190503}.

 \bibitem{Sangouard2008}
     Sangouard N., Simon C., Zhao B., Chen Y-A, De Riedmatten H., Pan J-W and Gisin N.
     Phys. Rev. A{77}{2008}{062301}.

\bibitem{Wang2012}
       Wang T. J., Song S. Y., Long G. L.
       Phys. Rev. A{85}{2012}{062311}.
       
       \bibitem{Zhao2003}
           Zhao Z., Yang T., Chen Y-A, Zhang A-N and Pan J-W
          Phys. Rev. Lett.{90}{2003}{207901}.


\bibitem{Tavis1968}
Tavis M. and Cummings F. W.
Phys. Rev.{170}{1968}{379}.

 \bibitem{Jaynes1963}
Jaynes E. T. and Cummings F. W.
Proc. IEEE{51}{1963}{89}.

   \bibitem{weinfurter1994}
  Weinfurter H.
    Europhys. Lett.{25}{1994}{559}.


   \bibitem{Liao2011}
Liao Q. H., Fang G. Y., Wang Y. Y., Ahmad M. A. and Liu S.
   Eur. Phys. J. D{61}{2011}{475}.

 \bibitem{Ghasemi2017}
Ghasemi M., Tavassoly M. K. and Nourmandipour A.
    Eur. Phys. J. Plus{132}{2017}{531}.

  \bibitem{Pakniat2017}
    Pakniat R., Tavassoly M. K. and Zandi M. H.
     Opt. Commun.{382}{2017}{381}.

\bibitem{Zukowski1993}
    Zukowski M., Zeilinger A., Horne M. A. and Ekert A. K.
    Phys. Rev. Lett.{71}{1993}{4287}.

\bibitem{Ian2016}
 Ian H.
    Europhys. Lett.{114}{2016}{50005}.

\bibitem{Lopez2017}
    L{\'o}pez C. E., Albarr{\'a}n-Arriagada F., Allende S. and Retamal J. C.
    Europhys. Lett.{120}{2017}{10003}.

\bibitem{Deng2017}
         Deng F. G., Ren B. C. and Li X. H.
         Sci. Bull.{62}{2017}{46}.

\bibitem{Qin2018}
         Qin W., Miranowicz A., Li P. B., L{\"u} X. Y., You J. Q. and Nori F.
         Phys. Rev Lett{120}{2018}{093601}.

\bibitem{Xie2016}
         Xie C. M., Liu Y. M., Chen J. L., Yin X. F. and Zhang Z. J.
          China-Phys. Mech. Astron{59}{2016}{100314}.

\bibitem{Ghasemi2016}
    Ghasemi M. and Tavassoly M. K.
     Eur. Phys. J. Plus{131}{2016}{297}.

\bibitem{Nourmandipour2016}
  Nourmandipour A. and Tavassoly M. K.
     Phys. Rev. A{94}{2016}{022339}.

\bibitem{nourmandipour2015}
    Nourmandipour A. and Tavassoly M. K.
     Eur. Phys. J. Plus{130}{2015}{148}.

\bibitem{xiao2008}
    Xiao L., Wang C., Zhang W., Huang Y., Peng J. and Long G.
     Phys. Rev. A{77}{2008}{042315}.

\bibitem{ai2008}
   Ai Q., Shi T., Long G. and Sun C.
     Phys. Rev. A{78}{2008}{022327}.


 \bibitem{Wootters1998}
   Wootters W. K.
  Phys. Rev. Lett.{80}{1998}{2245}.

 \bibitem{Obada2008}
     Obada A-S, Hessian H. A. and Mohamed A-A
      Opt. Commun.{281}{2008}{5189}.

\bibitem{Bashkirov2008}
 Bashkirov E. K., Rusakova M. S.
    Opt. Commun.{281}{2008}{4380}.

    \bibitem{Zhou2002}
  Zhou L., Song H. S., Li C.
    J. Opt. B: Quantum and Semiclassical Opt.{4}{2002}{425}.

     \bibitem{Deppe2008}
  Deppe F., Mariantoni M., Menzel E. P., Marx A., Saito S., Kakuyanagi K., Tanaka H., Meno T., Semba K., Takayanagi H., Solano E.
    Nature Phys.{4}{2008}{686}.

     \bibitem{Isenhower2010}
    Isenhower L., Urban E., Zhang X. L., Gill A. T., Henage T., Johnson T. A., Walker  T. G., Saffman M.
    Phys. Rev. Lett.{104}{2010}{010503}.

      \bibitem{Mohapatra2007}
    Mohapatra A. K., Jackson T. R., Adams C. S.
      Phys. Rev. Lett.{98}{2007}{113003}.


  \bibitem{James2007}
      James D. F. and Jerke J.
       Can. J. Phys.{85}{2007}{625}.

  \bibitem{Zheng2000}
      Zheng S-B and Guo G-C
       Phys. Rev. Lett.{85}{2000}{2392}.


\end{thebibliography}

\end{document}